\documentclass[usenatbib,useAMS]{mn2e}
\usepackage{amssymb}
\usepackage{graphicx}

\newcommand{\kms}{{{\rm \,km~s^{-1}}}}

\newcommand{\beq}{\begin{equation}}              
\newcommand{\eeq}{\end{equation}}                
\newcommand{\Mvir}{M_{\rm vir}}
\newcommand{\vcvir}{v_{c,\rm vir}}
\newcommand{\vcvirzero}{v_{c,\rm vir0}}
\newcommand{\vceff}{v_{c,\rm eff}}
\newcommand{\vcopt}{v_{c,\rm opt}}
\newcommand{\sigcent}{\sigma_{\rm cent}}
\newcommand{\sigmod}{\sigma_{\rm mod}}
\newcommand{\sigSIE}{\sigma_{\rm SIE}}
\newcommand{\sigSIEzero}{\sigma_{\rm{SIE}0}}
\newcommand{\cvir}{c_{\rm vir}}
\newcommand{\Fcool}{F_{\rm cool}}

\title
[Constraints on the velocity profiles of galaxies]
{Constraints on the velocity profiles of galaxies from strong lensing
statistics and semi-analytical modelling of galaxy formation}
\author[Chae et al.]{
Kyu-Hyun Chae,$^1$\thanks{chae@sejong.ac.kr}
Shude Mao$^2$
and Xi Kang$^3$ \\
$^1${Sejong University, Department of Astronomy and Space Sciences, 98
Gunja-dong, Gwangjin-Gu, Seoul 143-747, Republic of Korea} \\
$^2${University of Manchester, Jodrell Bank Observatory,
  Macclesfield, Cheshire SK11 9DL, UK} \\
$^3${Astrophysics Department, University of Oxford, Keble Road, Oxford
OX1 3RH, UK}
}
\date{
Accepted ........
Received .......;
in original form ......}

\pubyear{2006}

\begin{document}
\maketitle

\begin{abstract}
Semi-analytical models of galaxy formation can be used to predict the
evolution of the number density of early-type galaxies as a function of the
circular velocity at the virial radius, $\vcvir$.
Gravitational lensing probability and separation distribution on the other 
hand are sensitive to the velocity dispersion (or circular velocity) at about 
the effective radius. We adopt the singular isothermal ellipsoid (SIE) lens 
model to estimate the velocity dispersion at the effective radius. 
The velocity dispersion from strong lensing based on the SIE, $\sigSIE$ is 
then closely related to the observational central stellar velocity dispersion,
$\sigcent$; we have empirically $\sigSIE \approx \sigcent$. 
We use radio lenses from the Cosmic Lens All-Sky Survey and the PMN-NVSS 
Extragalactic Lens Survey to study how the velocity dispersions, $\sigSIE$,
are related to $\vcvir$; if  the galaxy were a singular isothermal sphere up 
to the virial radius, $\vcvir = \sqrt{2} \sigSIE$. 
When we include both the lensing probability and
separation distribution as our lensing constraints, we find
$\sigSIE /(200\kms) = 
 [(1.17_{-0.26}^{+0.40}) \vcvir/ (200\kms)]^{0.22^{+0.05}_{-0.04}}$
for $200\kms \la  \sigSIE \la 260\kms$; 
at $\sigSIE = 200\kms$, the ratio $\sqrt{2} \sigSIE / \vcvir$
is about $1.65^{+0.57}_{-0.37}$ (68\% CL) but decreases to 
$0.65_{-0.12}^{+0.15}$ (68\% CL) for $\sigSIE = 260\kms$.
These results are consistent with those of \citet{Sel02} obtained from
galaxy-galaxy weak lensing for galaxies of around $L_*$. However, our results
clearly suggest that the ratio must vary significantly as $\sigSIE$ is varied
and are marginally discrepant with the Seljak results at $\sigSIE = 260\kms$.
The scaling $\sigSIE \propto \vcvir^{0.22\pm 0.05}$ is broadly consistent with
those from galaxy occupation statistics studies and the most recent
galaxy-galaxy weak lensing study. We discuss briefly the implications of
our results for galaxy formations and structures.
These constraints can be significantly strengthened when larger
lens samples become available and the accuracy of semi-analytical model
predictions improves. 
\end{abstract}

\maketitle

\section{Introduction}

The $\Lambda$CDM structure formation model has been very successful
in explaining many observations, for example the cosmic microwave
background radiation (\citealt{Spe03, Spe06}), and the large-scale structures
(\citealt{Pea01, Teg04}).
In this model, the tiny quantum fluctuations in the early universe
are amplified due to gravitational instabilities, and eventually evolve
into highly nonlinear structures. The dissipationless cold
dark matter dominates the gravity, and determines the basic
abundance, the internal structure and formation history of the
nonlinear dark matter haloes. Such information can be reliably obtained using
$N$-body simulations and/or analytical models such as the Press-Schechter
formalism (e.g., \citealt{PS74, ST02}).
For example, the dark matter haloes follow approximately
the Navarro, Frenk and White (NFW) profile (\citealt{NFW97, Nav04}),
and their shapes can be approximated by tri-axial ellipsoidal models 
(\citealt{JS02}). Their time evolution can be studied numerically, which is 
also matched well by the analytical extended Press-Schechter formalism
(e.g., \citealt{Bon91, LC94, ST02}),

On the other hand, the visible galaxies
we see today are formed by the condensation of baryons within dark
matter haloes due to radiative processes. When the baryons sink toward
the centre of dark haloes and become self-gravitating, star formation,
active galactic nuclei and feedback processes will occur. Unfortunately,
these phenomena are not well-understood.
For clusters of galaxies, baryonic cooling is not very important, so
their mass profiles still approximately follow the predictions from
$N$-body simulations, as seen from kinematic (e.g., \citealt{vdM00}), 
lensing (e.g., \citealt{Com06}) and X-ray (\citealt{VF06}) studies of 
clusters of galaxies. On the scale of galaxies, cooling is  important, and 
the central profiles of galaxies are likely to be significantly affected by 
baryonic processes. Hydrodynamical high-resolution simulations can now simulate
regions of the universe or individual galaxies (e.g., \citealt{Mez03}),
but they cannot yet resolve the internal structure of galaxies and at the same
time simulate large enough volume to be statistically representative (and 
realistic). Nevertheless, rapid progress has been made, including more 
realistic modelling of the multi-phase interstellar medium (\citealt{SH03}).
An alternative approach is to use semi-analytical studies which
can incorporate many physical processes in an intuitive way
(\citealt{Kau99,SP99,Col00,Kan05,Cro06}).
These models can reproduce the luminosity function and correlation
function of galaxies reasonably well. However,
although these models can provide the number
of galaxies as a function of circular velocity at the virial radius,
they cannot yet reliably predict the central dynamical properties
of galaxies, such as the central velocity dispersions.

The central velocity dispersions (or the velocity dispersions in the optical 
regions) of galaxies are precisely the information needed for the studies of 
gravitational lensing, as the lensing probability is proportional to $\sigma^4$
while the image separation is proportional to $\sigma^2$. Earlier studies
(e.g., \citealt{NW88, Koc95}) used singular isothermal spheres to constrain
variants of CDM models. In this scenario, the velocity dispersion
is simply related to the circular velocity at the
virial radius by $v_c =\sqrt{2} \sigma$. However, this assumption is
likely to be invalid for two reasons. The NFW profile for haloes
does not give rise to a flat rotation curve. In fact, the value of its
peak is about 20 per cent higher than that at the virial radius
for a concentration parameter ($\approx 10$) appropriate for Milky-Way
sized haloes. Second, the baryons settled at the centre may further increase
the central rotations and velocity dispersions.
As the assembly of galaxies is not yet well understood,
how the velocity dispersion in the optical region
relates to the virial circular velocity is therefore a parameter
we want to extract from observations. In this paper, we propose to
use strong gravitational lensing to constrain this key parameter.

At the time of this writing no study has made use of the predictions
of semi-analytical studies of galaxy abundances of different types.
Several studies have used gravitational lensing to constrain the
evolution of galaxies (\citealt{Mao91, MK94,Ofe03, CM03}). 
In this paper, we will use the predicted abundance of ellipticals  from
semi-analytical models as a function of virial circular velocity. 
We then adopt a simple form to parameterise the relation between the
velocity dispersion at the optical radius and the circular velocity at the 
virial radius. We use radio lenses from the Cosmic Lens All-Sky Survey 
(CLASS; \citealt{Mye03, Bro03})
and the PMN-NVSS Extragalactic Lens Survey (PANELS; \citealt{Win01b})
to constrain this factor empirically. This in turn will provide an important
way of understanding the baryonic effects on the central properties of
galaxies. Our approach is independent of the galaxy-galaxy weak
lensing studies of the same factor (\citealt{Sel02}), who
found that for both early-type and late-type galaxies around $L_*$,
their peak velocities are about a factor from 1.7 to 1.8 (with an uncertainty
of about 20\%) of the circular velocity at the virial radius.
The structure of this paper is as follows.
In \S2, we outline our method and the data to be used. In \S3, we
present the results of our analyses. Finally, in \S4 we discuss
the implications of our results for galaxy formation along with possible
sources of systematic errors and give the main conclusions of our study.

\section{Model, Method and Data}

In \S 2.1 we introduce velocity functions of galaxies starting from the
Schechter luminosity function (\citealt{Sch76}) of galaxies. We also adopt a 
model for the relation between the velocity dispersion arising from strong 
lensing and the virial circular velocity predicted from semi-analytical 
modelling of galaxy formation (\S 2.1). In \S 2.2 we briefly describe
our adopted semi-analytical model (SAM) of galaxy formation and present
simulated data to be used in this work. We also outline our formalism 
of strong lensing statistics to constrain the radial velocity profile of 
galaxies based on the SAM circular velocity function and strong lensing data.

\subsection{Velocity functions of galaxies}

The number density of galaxies as a function of luminosity is described by the
Schechter luminosity function (LF) $\phi_{\rm L}$ given by
\beq
dn = \phi_{\rm L} (L) dL = \phi_*
  \left(\frac{L}{L_{*}}\right)^{\alpha_{\rm L}}
   \exp\left(-\frac{L}{L_{*}}\right)
    \frac{dL}{L_{*}}.
\label{LF}
\eeq
We assume an effective power-law relation between the luminosity ($L$) and the
velocity dispersion ($\sigma$) for the scale under consideration of galaxies 
given by\footnote{In reality the relation between $L$ and $\sigma$ does not 
follow an exact power-law relation but suffers from a significant scatter. 
Thus, the power-law fitting function between $L$ and $\sigma$, namely the 
Faber-Jackson relation (\citealt{FJ76}), may not be the same as this
effective relation (see \citealt{She03}).}
\beq
\frac{L}{L_*} =   \left(\frac{\sigma}{\sigma_*}\right)^{\beta_{\rm VD}}.
\label{LVD}
\eeq
Then the number density of galaxies as a function of velocity dispersion can
be described by the velocity dispersion function (VDF) $\phi_{\rm VD}$ given by
\beq
dn = \phi_{\rm VD} (\sigma) d\sigma = \phi_*
  \left(\frac{\sigma}{\sigma_*}\right)^{\alpha_{\rm VD}-1}
   \exp\left[-\left(\frac{\sigma}{\sigma_*}\right)^{\beta_{\rm VD}}\right]
   \beta_{\rm VD} \frac{d\sigma}{\sigma_*},
\label{VDF}
\eeq
where $\alpha_{\rm VD} = (\alpha_{\rm L}+1)\beta_{\rm VD}$.

In this work we shall study how the velocity dispersion in the optical region
and the inner halo constrained by strong lensing statistics is related to 
the circular velocity at the virial radius ($\vcvir$) predicted by an 
up-to-date semi-analytical model of galaxy formation. 
Notice that the velocity dispersion constrained by strong lensing
depends on the adopted lens model (e.g., \citealt{Cha03}). We shall denote
the lens model-dependent velocity dispersion by $\sigmod$. Parameter $\sigmod$ 
corresponds to the velocity dispersion  approximately at the scale of the 
optical region and the inner part of the halo because the observed 
galactic-scale lens images are formed at around the effective radii. 
In particular, although a lens model may imply a mass distribution from the 
galactic centre to infinity, the mass distribution well outside the optical
region is an extrapolation that is not sensitive to strong lensing.
We expect that the velocity dispersion $\sigmod$ is closely related to the 
central stellar velocity dispersion, $\sigcent$ that is observable
in the aperture-limited spectroscopic observations. In this work we shall 
assume a constant proportionality between the two parameters; we write 
\beq
\sigmod = \eta \sigcent
\label{modcent}
\eeq
(see \S 4 for a discussion on the possibilities of varying $\eta$).   

\begin{table}
\caption{Summary of velocity and related parameters
\label{complete}}
\begin{tabular}{lll}
\hline
 parameter  &  meaning & relevant scale  \\
\hline
 $\vcvir$  & circular velocity at the virial radius  & $r_{\rm vir}$  \\
 $\vceff$  & circular velocity at the effective radius & $r_{\rm eff}$  \\
 $\sigmod$ & lens model velocity dispersion  & $\sim r_{\rm eff}$ \\
 ($\sigSIE$) & (SIE model velocity dispersion)  &  \\
 $\sigcent$ & observed central velocity dispersion  &  $\la r_{\rm eff}$  \\
 $\sqrt{2}\sigSIE$ & SIE model circular velocity  & $\sim r_{\rm eff}$ \\
 $\eta$  &  $\equiv \sigmod/\sigcent$  &     \\
 $f$  &  $\equiv \sigmod/\vcvir$  &      \\
\hline
\end{tabular}

Notes. $r_{\rm vir}=$ virial radius.  $r_{\rm eff}=$ effective radius.
\end{table}

Hereafter for the sake of simplicity parameters $\sigma$ and $v_c$ shall refer
respectively to $\sigmod$ and $\vcvir$ unless specified otherwise. We 
parameterise the relationship between the two parameters $\sigma$ and $v_c$ by
\beq
\sigma = f(v_c) v_c.
\label{fdef}
\eeq
 This is reasonable as on
the cluster scale, the baryonic effect may be small, while the baryonic
condensation within galactic-sized halos may significantly boost the velocity 
dispersion in the optical region relative to the virial circular velocity.
We adopt a model given by
\beq
 f(v_c) = f_* \left(\frac{v_c}{v_{c*}}\right)^\mu,
\label{fmod}
\eeq
where $\mu=0$ corresponds to the case where the velocity dispersion is related
to the virial circular velocity by a constant proportionality. For the above 
model of equation~(\ref{fmod}) the virial circular velocity function (CVF)
 $\phi_{\rm CV}$ takes the following form\footnote{This particular form is
the same functional form as the VDF given by equation~(\ref{VDF}). This is the
direct consequence of adopting the particular model of equation~(\ref{fmod}).}
\beq
dn = \phi_{\rm CV} (v_c) dv_c = \phi_*
  \left(\frac{v_c}{v_{c*}}\right)^{\alpha_{\rm CV}-1}
   \exp\left[-\left(\frac{v_c}{v_{c*}}\right)^{\beta_{\rm CV}}\right]
   \beta_{\rm CV} \frac{dv_c}{v_{c*}},
\label{CVF}
\eeq
where we have the following relations:
\beq
\alpha_{\rm CV} = \alpha_{\rm VD} (\mu+1), \hspace{0.3cm}
\beta_{\rm CV} = \beta_{\rm VD} (\mu+1),\hspace{0.3cm} {\rm and}
\hspace{0.2cm} \sigma_* = f_* v_{c*}.
\label{pararels}
\eeq
The above CVF (equation~\ref{CVF}) implies an effective power-law relation 
between the luminosity and the virial circular velocity of galaxies given by
\beq
\frac{L}{L_*} =   \left(\frac{v_c}{v_{c*}}\right)^{\beta_{\rm CV}}.
\label{LCV}
\eeq
The velocity and related parameters introduced above are summarised in Table~1.

Finally, a galaxy of $v_{c*}$ may
not correspond to a typical $L_*$ galaxy. Suppose an $L_*$ galaxy has a
velocity dispersion of $\sigma_0$. Then the value of $f$ (equation~\ref{fmod}) 
for $\sigma_0$ is given by
\beq
f_0 = f_* \left( \frac{\sigma_0}{\sigma_*} \right)^{\mu/(\mu+1)}.
\label{f0star}
\eeq
We shall use parameter $f_0$ rather than $f_*$ for a chosen value of
$\sigma_0$.

\subsection{Strong lensing statistics and semi-analytical model of galaxy
formation}

The statistical properties of strong lensing are determined by four
ingredients, namely the number density of potential lenses as a function of
redshift, the lens cross sections, the cosmology, and the magnification bias
depending on the potential source population (see, e.g., \citealt{Cha03}).
Most studies of strong lensing statistics have used the present-day galaxy
population given in the form of luminosity (or velocity dispersion) function
for the number density of potential lenses (see, e.g., \citealt{TOG, Fuk92,
Koc96, Hel99, Cha03, Mit05}). In this paper, however, we adopt
the number density of galaxies from the recent semi-analytical model of 
\citet{Kan05}, which gives the number density of galaxies as a function
of redshift and the circular velocity at the virial radius.

The galaxy catalogue constructed by \citet{Kan05} used a
high-resolution N-body simulation by \citet{JS02}.
The simulation follows the evolution of $512^{3}$ particles in a cosmological
box of $100 h^{-1} {\rm Mpc}$ with the standard concordance cosmological model 
($\Omega_{\rm m,0} =0.3, \Omega_{\Lambda, 0} =0.7, \sigma_{8}=0.9, h=0.7$).
\citet{Kan05} adopted reasonable treatment of physical processes in
predicting galaxy population such as cooling, star formation rate and
energy feedbacks from supernova, and dust extinction. The parameters
in the model were normalized by the local galaxy luminosity functions.
The readers are refereed to that paper for more detail.

\begin{table*}
\caption{Distribution of the number of early-type and late-type galaxies as
a function of virial circular velocity predicted by the \citet{Kan05}
semi-analytical model of galaxy-formation. Each value given here is the
number of early-type (late-type) galaxies in a velocity bin of width
$25 \kms$ contained within a total comoving volume of
$10\times 10^6$~$h^{-3}$~Mpc$^3$ distributed for the redshift range of
$0.3 \la z \la 1$.
\label{complete}}
\begin{tabular}{clll}
\hline
 $i$ & $\vcvir(i)$ ($\kms$) & $\Delta N^{\rm (early)}(i)$
 & $\Delta N^{\rm (late)}(i)$ \\
 (bin \#)  &  (mean virial circular velocity) & (\# of early-type galaxies)
           & (\# of late-type galaxies)  \\
\hline
1  &  112.5  & 30321  &  148585  \\
2  & 137.5   & 12854  &  73322   \\
3  &  162.5  &  5816  &  40381   \\
4  &  187.5  &  2789  &  22903   \\
5  &  212.5  &  1719  &  14661   \\
6  &  237.5  &  1377  &  8802    \\
7  &  262.5  &  1266  &  5899   \\
8  &  287.5  &  830   &  3808   \\
9  &  312.5  &  763   &  2357   \\
10  &  337.5  &  688  &  1484   \\
11  &  375.  &  532   &  830   \\
12  &  425.  &  295   &  442   \\
13  &  475.   & 186   &   260   \\
14  &  525.   & 145   &  120   \\
15  &  575.   & 108   &  72   \\
\hline
\end{tabular}

\end{table*}

\begin{table*}
\caption{Strongly lensed systems from the CLASS (\citealt{Bro03}) and the
PANELS southern sky (\citealt{Win00,Win01a,Win01b,Win02a,Win02b}) radio 
surveys (taken from \citealt{Cha05}). The systems under the
`CLASS statistical' survey are members of the well-defined CLASS statistical
sample of 8958 radio sources (\citealt{Bro03, Cha03}).
\label{survey}}
\begin{tabular}{lllllll}
\hline
Source &  Survey & Source & Lens  & Image Separation & Image & Lens Type \\
    &   & Redshift & Redshift & (arcsec) &  Multiplicity &  \\
\hline
 B0128+437 & CLASS  & 3.124 & 1.145 & 0.54 & 4 & unknown  \\
 J0134$-$0931 & PANELS  &  2.225 & 0.7645 & 0.681 & 2+4 & 2Gs  \\
 B0218+357  & CLASS statistical &  0.96  & 0.68  & 0.334  & 2  & spiral \\
 B0414+054  & CLASS  & 2.62   & 0.958 & 2.09   & 4  & early-type \\
 B0445+123 & CLASS statistical & --- & 0.558 & 1.33 & 2 & early-type \\
 B0631+519 & CLASS statistical & --- & 0.620 & 1.16 & 2 & early-type \\
 B0712+472 & CLASS statistical & 1.34 & 0.41 & 1.27 & 4 & early-type \\
 B0739+366 & CLASS   & ---   &  ---    & 0.54   & 2  & unknown  \\
 B0850+054  & CLASS statistical &  ---   & 0.588 & 0.68   & 2  & spiral \\
 B1030+074  & CLASS   &  1.535 & 0.599 & 1.56   & 2  & early-type  \\
 B1127+385 & CLASS   &   ---  &  ---  & 0.70   & 2  & spiral  \\
 B1152+199 & CLASS statistical & 1.019 & 0.439 & 1.56 & 2 & unknown \\
 B1359+154  & CLASS statistical &  3.235 & ---  & 1.65   & 6  &  3Gs \\
 B1422+231 & CLASS statistical & 3.62 & 0.34 & 1.28 & 4 & early-type \\
 B1555+375 & CLASS   &  ---   &  ---   & 0.42  & 4  & unknown   \\
 B1600+434 & CLASS   & 1.57   & 0.415 & 1.39   & 2  & spiral  \\
 B1608+656  & CLASS statistical &  1.39  & 0.64  & 2.08   & 4  &  2Gs \\
 J1632$-$0033 & PANELS  & 3.42  &  1  & 1.47   & 2  & early-type  \\
 J1838$-$3427 & PANELS & 2.78  & 0.36 & 1.0   & 2  & early-type  \\
 B1933+503 & CLASS statistical & 2.62 & 0.755 & 1.17 & 4 & early-type \\
 B1938+666 & CLASS   & $\ga 1.8$ & 0.881 & 0.93 & ring & early-type \\
 J2004$-$1349 & PANELS & --- &  ---  &  1.13  &  2  & spiral  \\
 B2045+265  & CLASS statistical & ---   & 0.867  & 1.86   & 4  & puzzling \\
 B2108+213  & CLASS   &  --- & 0.365  & 4.55   & 2 or 3 & 2Gs+cluster \\
 B2114+022  & CLASS statistical & ---  & 0.32/0.59 & 2.57  & 2 & 2Gs \\
 B2319+051  & CLASS statistical & --- & 0.624/0.588 & 1.36 & 2
       & early-type \\
\hline
\end{tabular}

\end{table*}

The \citet{Kan05} SAM reproduces the luminosity function of galaxies
reasonably well for the range of luminosity suitable for strong lensing
studies. The semi-analytical studies provide a simple way of dividing galaxies
into two populations, namely the early-type (ellipticals and S0's) population
and the late-type population based on the correlation between the B-band
bulge-to-disk ratio and the Hubble type of the galaxy
(\citealt{Sd86}). We use only the early-type population for
this work because the radio lens sample we shall use contains only a small
 number of spiral lensing galaxies. Table~2
shows the numbers of early-type and late-type galaxies in velocity bins
contained in $10 \times 10^6$ $h^{-3}$~Mpc$^3$ volume distributed for the
redshift range of $0.3 \la z \la 1$.

While the SAM data can give a virial circular velocity function,
strong lensing is not sensitive to the virial circular velocity but the
velocity dispersion in the optical region and the inner halo. Hence to proceed
in lensing computations using the number density of early-type galaxies from
the SAM we must relate the virial circular velocity to the velocity
dispersion. We proceed as follows. We adopt the VDF $\phi_{\rm VD}(\sigma)$
(equation~\ref{VDF}) for lensing computations  and the CVF $\phi_{\rm CV}(v_c)$
(equation~\ref{CVF})
for describing the SAM data assuming that they are related by the power-law
model given by equation~(\ref{fmod}). We then fit lensing data and the SAM data
simultaneously. To fit the VDF to lensing data we use a maximum likelihood
method based on those recently used by \citet{Cha02}, \citet{Cha03} and
\citet{Cha05}. The likelihood for lensing is defined by
\beq
\ln \mathcal{L} =
          \left( \sum_{j=1}^{N_{\rm IS}} w_j \ln \delta p_{\rm IS}(j) \right)
                + \left( \sum_{k=1}^{N_{\rm U}} \ln [1 - p(k)]
                + \sum_{l=1}^{N_{\rm L}} \ln \delta p(l) \right),
\label{Lhood}
\eeq
where the first term is the likelihood due to the relative image separation
probabilities of lensed sources as defined by \citet{Cha05} and the terms in
the second parenthesis are the likelihood due to the CLASS statistical sample
(\citealt{Bro03}) as defined by  \citet{Cha02} and \citet{Cha03};
parameters $w_k$ are the weight factors, $\delta p_{\rm IS}(j)$
(equation~4 of \citealt{Cha05}) are the relative image separation
probabilities for lensed sources with well-defined image separations
(Table~3; see also Table~1 of \citealt{Cha05}),
$p(k)$ (equation~40 of \citealt{Cha03}) are the total lensing probabilities
for the unlensed sources in the CLASS statistical sample (see \S 3.1 and
\S 3.2 of \citealt{Cha03}), and $\delta p(l)$ (equations~29, 38, or 39 of
\citealt{Cha03})
are the differential lensing probabilities for the lensed sources in the CLASS
statistical sample (Table~3; see also Table~1 of \citealt{Cha03}).
Notice that the above lensing probabilities are calculated assuming the 
singular isothermal ellipsoid (SIE) lens model whose projected surface density
 is given by (\citealt{Cha03})
\beq
\Sigma_{\rm SIE}(x,y)= \frac{\sigSIE^2}{2 G} 
 \frac{\sqrt{e} \lambda(e)}{\sqrt{x^2+e^2 y^2}},
\label{SIE}
\eeq
where $\sigSIE$ is the model velocity dispersion, $e$ is the axis 
ratio, and $\lambda(e)$ is the `dynamical normalisation' factor 
(\S~2.1.1 of \citealt{Cha03}). We assume that galaxies are not biased toward 
either oblate or prolate shape and have a mean projected ellipticity of 
$\epsilon (\equiv 1-e) \la 0.45$ so that $\lambda(e) \approx 1$.
Notice that the shape and characteristic velocity dispersion of the VDF are
most directly constrained by the image separation distribution. The absolute
 lensing probability of the CLASS statistical sample is also sensitive to
the VDF as the lensing probability is proportional to $\phi_* \sigma_*^4$.
The chi-squared for lensing is then defined by
\beq
\chi^2_{\rm lens} = - 2 \ln \mathcal{L}.
\label{chilens}
\eeq
To fit the CVF (equation~\ref{CVF}) to the SAM data (Table~2) we define the 
following chi-squared function
\beq
\chi^2_{\rm SAM} = \sum_{i=1}^{N_{\rm bin}}
   \left(\frac{\phi_{\rm CV}(v_c(i)) \Delta v_c \Delta V - \Delta N (i)}
           {\sqrt{\Delta N (i)}} \right)^2,
\label{chiSAM}
\eeq
where $\Delta v_c = 25 \kms$ and $\Delta V = 10^7$ $h^{-3}$~Mpc$^3$.
Finally, the total chi-squared function is defined to be
\beq
\chi^2_{\rm tot} = \chi^2_{\rm lens} + \chi^2_{\rm SAM}.
\label{chitot}
\eeq

\section{Results}

\begin{figure}
\begin{center}
\setlength{\unitlength}{1cm}
\begin{picture}(10,10)(0,0)
\put(-1.,-2.){\includegraphics{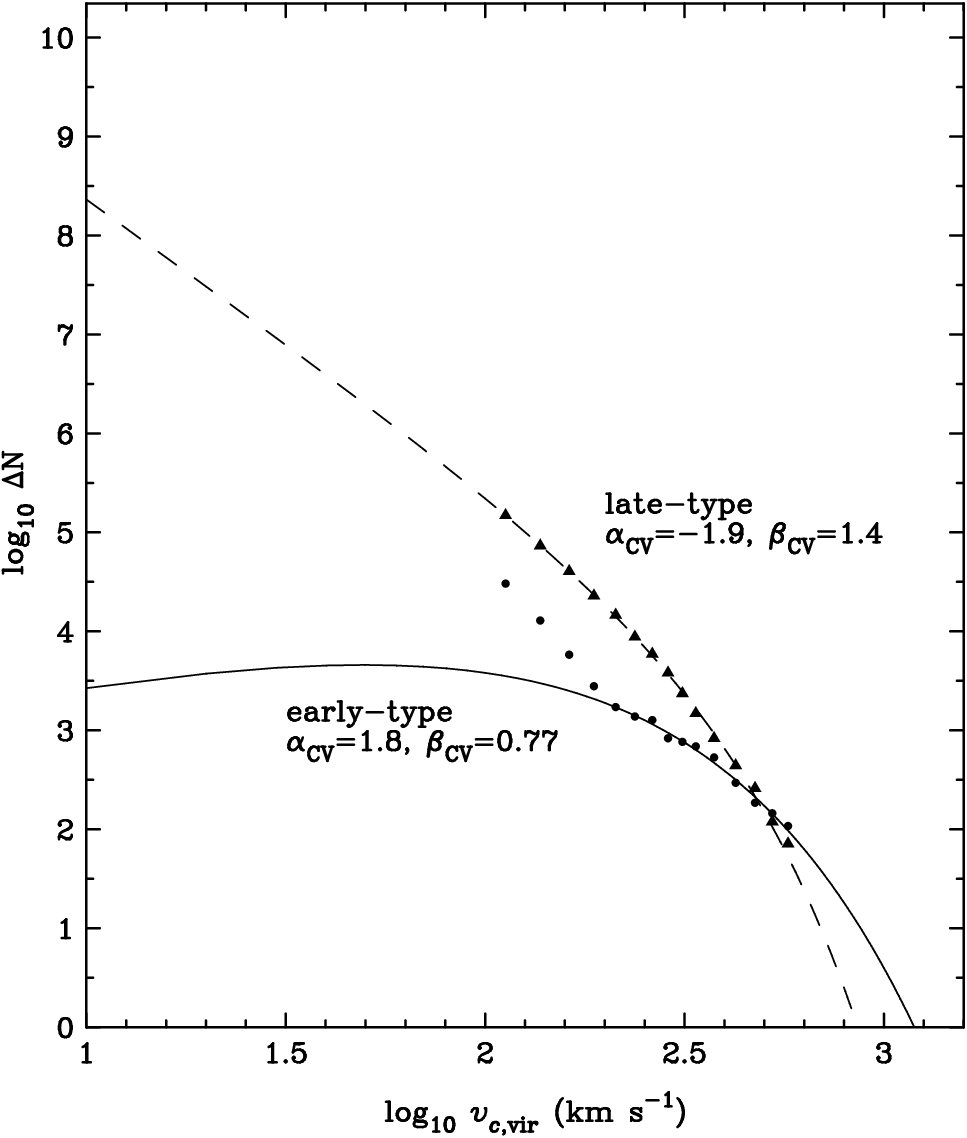}}
\end{picture}
\caption{
The numbers of early-type galaxies and late-type galaxies (Table~2) as a
function of circular velocity at the virial radius predicted by the 
\citet{Kan05} semi-analytical model of galaxy formation.
The circular velocity functions fitted to the data using equation~(\ref{CVF}) 
are also shown. We have used Poisson errors for the fitting,
which are however too small to display, and ignored the first four data points
of the early-type population (see the texts). The results are for
$10 \times 10^6$~h$^{-3}$~Mpc$^3$ comoving volume nearly uniformly distributed
over the redshift range of $0.3 \la z \la 1$. The velocity bin size is
$25 \kms$.
}
\label{fig:cvf}
\end{center}
\end{figure}

\begin{figure*}
\begin{center}
\setlength{\unitlength}{1cm}
\begin{picture}(8,8)(0,0)
\put(-6.,12.){\includegraphics{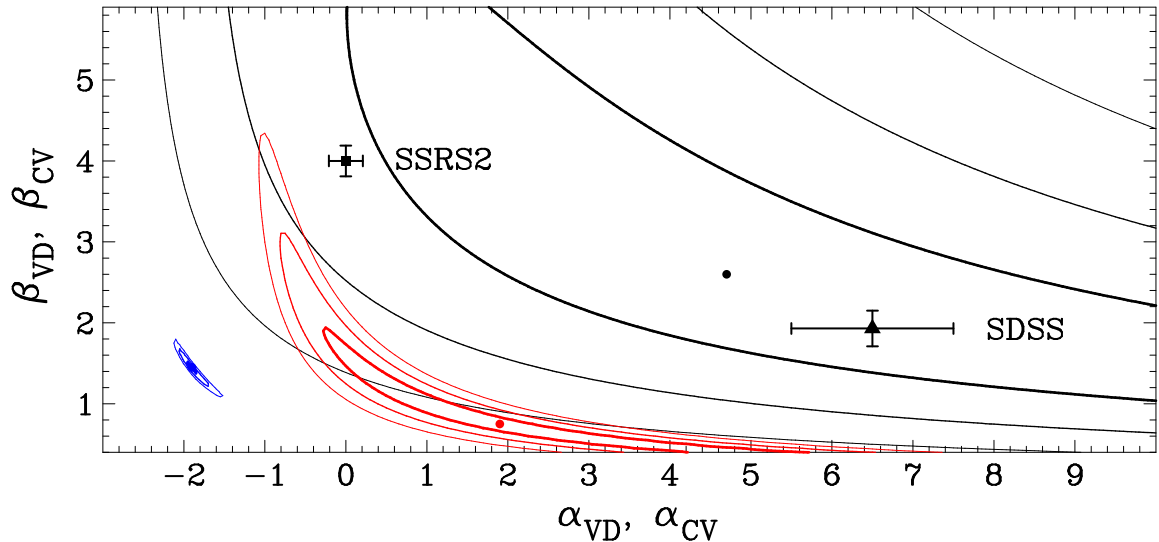}}
\end{picture}
\caption{
Constraints on the velocity functions of galaxies. The red and blue contours
are respectively the confidence limits (CLs) on the CVFs (equation~\ref{CVF}) 
for the early-type and the late-type populations as obtained from the 
\citet{Kan05} SAM simulated data.  The black contours are the CLs on the VDF
(equation~\ref{VDF}) for the early-type population as obtained by \citet{Cha05}
from the image separation distribution of gravitational lenses. The three 
contours on each function correspond respectively to $1\sigma$, $2\sigma$, and
$3\sigma$. The filled triangle represents the SDSS VDF as obtained by 
\citet{She03} from the measured central velocity dispersions of
early-type galaxies. The filled square represents the SSRS2 VDF as obtained
by \citet{Cha03} from the \citet{Mar98} early-type LF and a measured
Faber-Jackson relation.
}
\label{fig:allvf}
\end{center}
\end{figure*}

We first examine the simulated data from the \citet{Kan05} SAM.
Fig.\ \ref{fig:cvf} shows the SAM data points (Table~2) along with the fitted 
CVFs (equation~\ref{CVF}) through the chi-squared function given by 
equation~(\ref{chiSAM}). Notice that the lowest 4 data
points for the early-type population are not included in the fitting because
they do significantly deviate from the behaviour of the rest of the data
points. We shall not use these 4 data points in our study of the velocity
profile of early-type galaxies (see below and \S 4). 
Fig.\ \ref{fig:allvf} shows the constraints on 
the shapes of the CVFs and compares with the
shape of the VDF of the early-type population obtained by \citet{Cha05}.
Notice that the CVF shape is significantly different from the VDF shape for
the early-type population. This means that the velocity dispersion cannot be
related to the virial circular velocity by a constant proportionality
(i.e.\ the case $\mu =0$ in equation~\ref{fmod}).

We simultaneously fit the VDF (equation~\ref{VDF}) and the CVF 
(equation~\ref{CVF}) for the early-type galaxy population assuming the relation
between the two given by equations~(\ref{fdef}) and (\ref{fmod}) to the lensing
data and the SAM data by minimising the total chi-squared function 
$\chi^2_{\rm tot}$ given by equation~(\ref{chitot}). We do not use the first 4
data points ($i=1$ to 4 in Table~2) for the early-type population from the SAM.
These 4 data points do not fit into the model CVF (equation~\ref{CVF}) while 
all the rest do (Fig.\ \ref{fig:cvf}).  These data points might be an artifact
due to the currently imperfect SAM of \citet{Kan05} (see \S4). Perhaps, more
importantly, lowest velocity galaxies may not be effective for strong lensing
for the following two reasons: the lensing cross-section scales as $\sigSIE^4$
and their inner mass profiles may be too shallow (e.g., \citealt{deB05}, see,
however, \citealt{Swa03}).
So simply ignoring them would make small errors in our study.
Nevertheless, we must bear in mind that any of our derived results for
$\vcvir \la 190 \kms$ will be an extrapolation and the strict range of
validity for our analyses is 
$200 \kms \la \vcvir \la 600\kms$ (see Table~2).

\begin{figure}
\begin{center}
\setlength{\unitlength}{1cm}
\begin{picture}(10,10)(0,0)
\put(-1.,-2.){\includegraphics{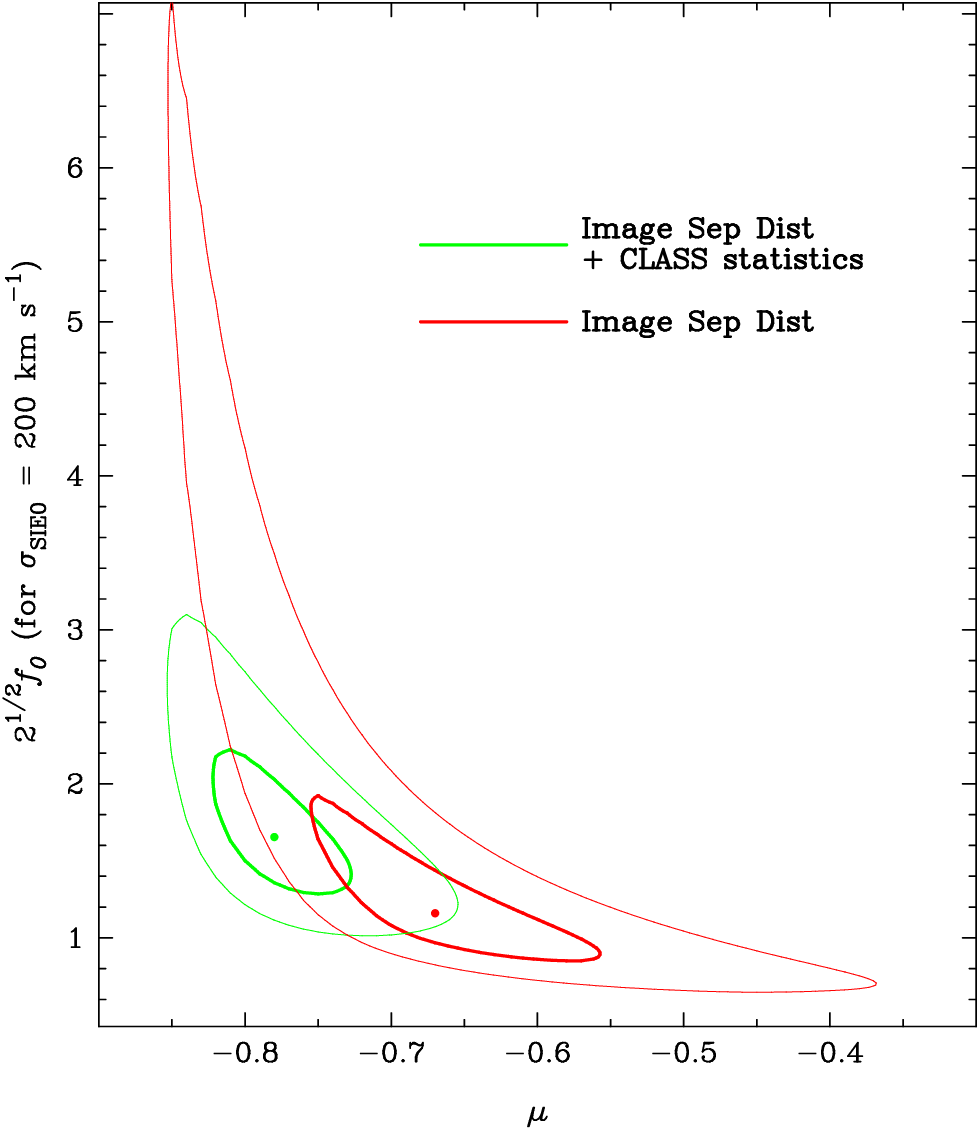}}
\end{picture}
\caption{
Likelihood contours in the plane of $\mu$ and $\sqrt{2} f_0$ based on
strong lensing and semi-analytical model of galaxy formation. Here parameter 
$\mu$ is the power-law index in the model $f(\vcvir)$ (equation~\ref{fmod}) for
the ratio of the velocity dispersion (in the optical region and the inner
halo) to the virial circular velocity and parameter $f_0$ is the value of the
ratio at the fiducial velocity dispersion of $\sigSIEzero = 200 \kms$.
The red contours are based only on the image separation distribution of
radio lenses. The green contours are based on both the image separation
distribution of radio lenses and all the lensing properties (including
the lensing rate) of the CLASS statistical sample. The two contours for each
case correspond respectively to the 68\% and 95\% confidence limits.
}
\label{fig:mu-f0}
\end{center}
\end{figure}

\begin{figure*}
\begin{center}
\setlength{\unitlength}{1cm}
\begin{picture}(11,11)(0,0)
\put(-4.5,13.){\includegraphics{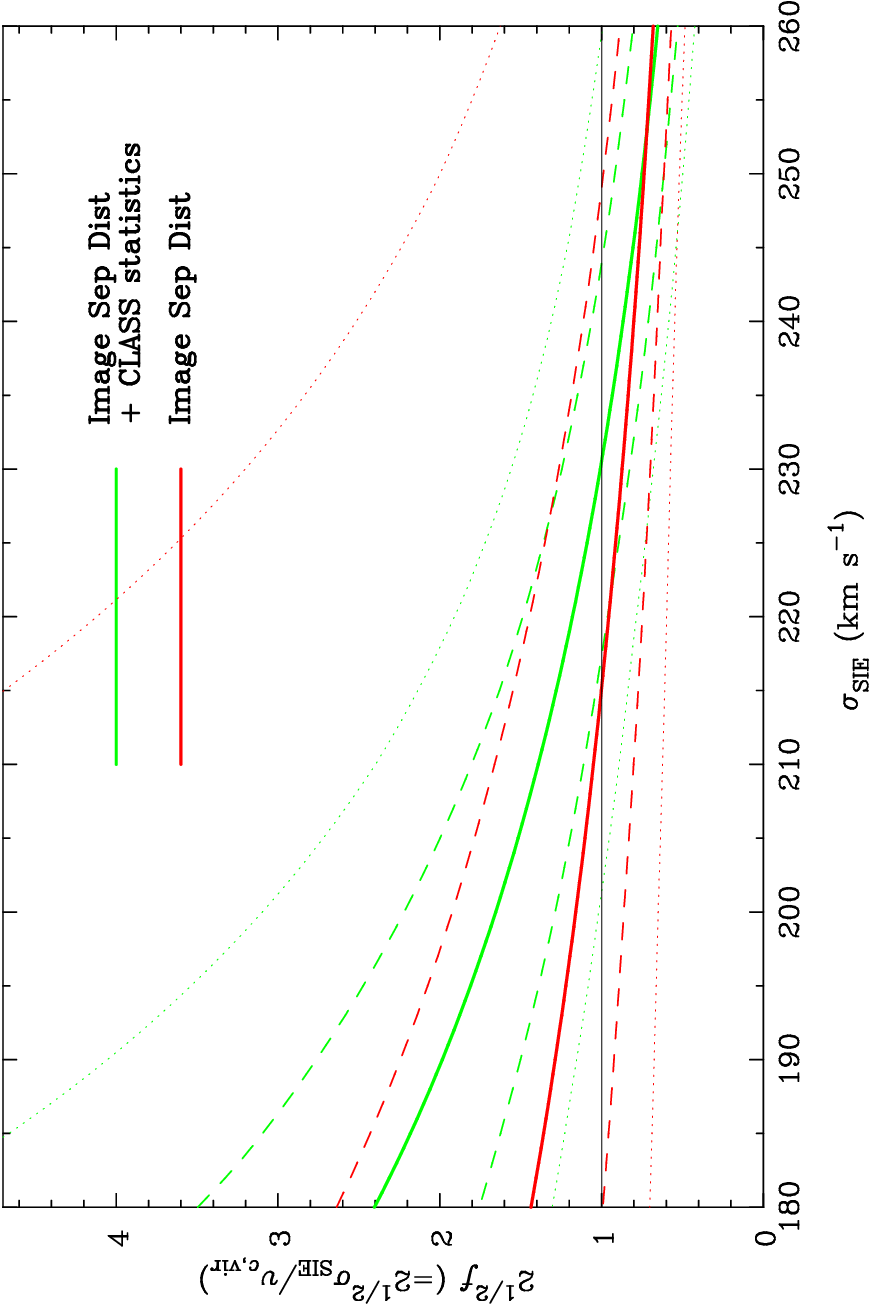}}
\end{picture}
\caption{
The behaviour of $\sqrt{2} f$ ($=\sqrt{2} \sigSIE/\vcvir$) as a function of
$\sigSIE$. Here parameter $\sigSIE$ is the velocity dispersion (in the optical
region and the inner halo) implied by strong lensing assuming the singular
isothermal ellipsoid model of \citet{Cha03} and
parameter $\vcvir$ is the circular velocity at the virial
radius of the halo predicted by the \citet{Kan05} semi-analytical model
of galaxy formation. For our adopted model of the singular isothermal
ellipsoid, the factor $\sqrt{2} \sigSIE$ corresponds to the circular velocity
in the optical region and the inner halo $\vceff$. The red line is based
only on the image separation distribution of radio lenses. The green line
is based on both the image separation distribution of radio lenses and all
the lensing properties (including the lensing rate) of the CLASS statistical
sample. The dashed and dotted lines represent respectively the 68\% and 95\%
confidence limits. Notice that the uncertainty in the ratio becomes larger
at smaller velocity for a given uncertainty of the velocity.
}
\label{fig:fv}
\end{center}
\end{figure*}

\begin{figure}
\begin{center}
\setlength{\unitlength}{1cm}
\begin{picture}(10,10)(0,0)
\put(-1.,-2.){\includegraphics{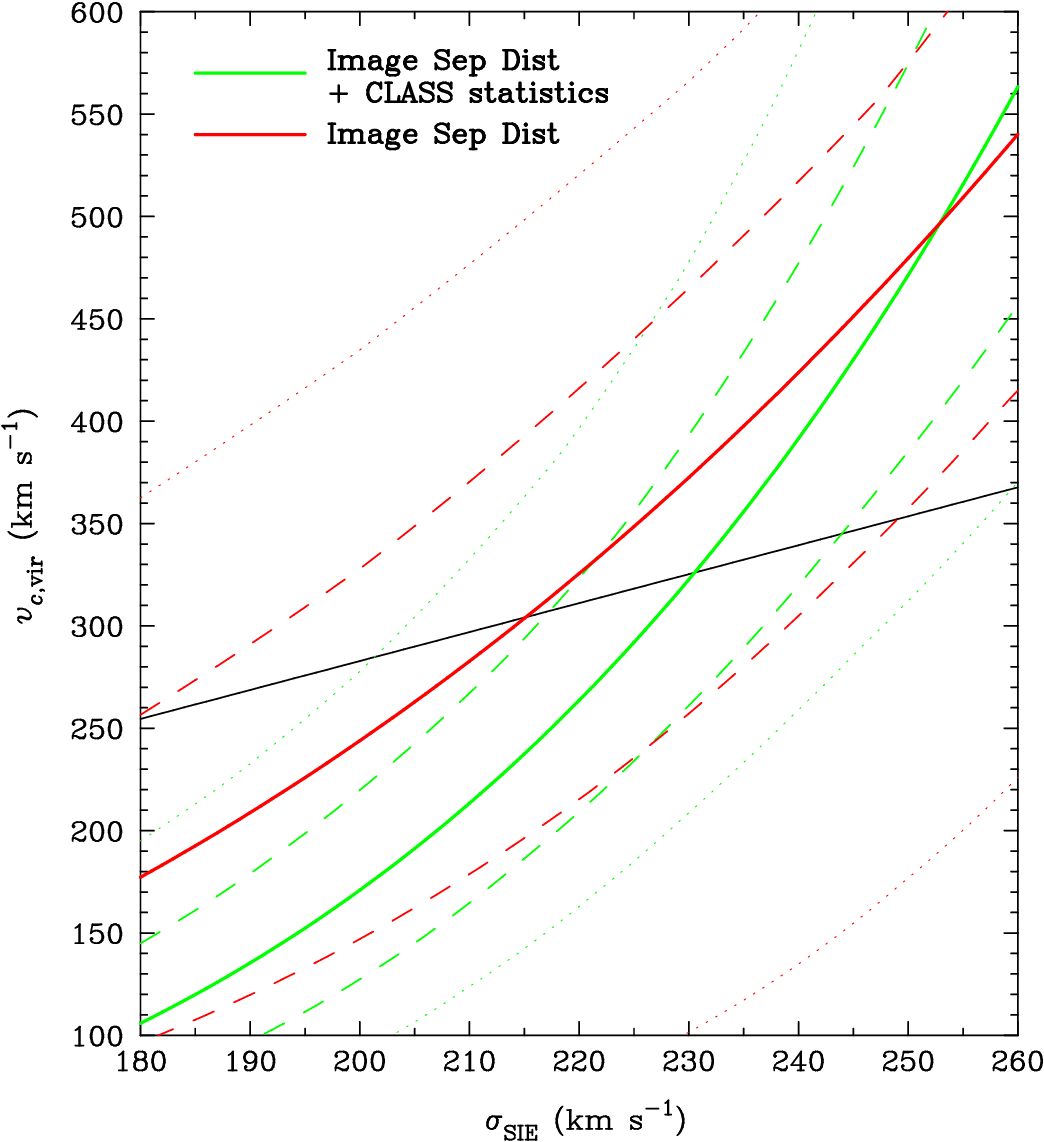}}
\end{picture}
\caption{
The behaviour of $\vcvir$ as a function of $\sigSIE$. Here
parameters $\vcvir$ and $\sigSIE$ are the same as in Fig.\ \ref{fig:fv}.
The red and green lines are also the same as in Fig.\ \ref{fig:fv}.
The dashed and dotted lines represent respectively the 68\% and 95\% confidence
limits. The solid black line represents $\vcvir = \sqrt{2} \sigSIE$.
Notice that the uncertainty in $\vcvir$  is nearly unchanged
 as $\sigSIE$ is varied.
}
\label{fig:sigma-vc}
\end{center}
\end{figure}

We consider first the case where we use both of the parts in 
equation~(\ref{Lhood}), namely both the relative image separation probabilities
 and the lensing properties of the CLASS statistical sample.
For the CLASS statistical sample (\citealt{Bro03, Cha03})
 the number of sources that are not strongly lensed is
$N_{\rm U} = 8945$. There are 13 strongly lensed sources in the CLASS
statistical sample as can be found in Table~3. However, the lenses for
 B0218+357 and  B0850+054 are spiral galaxies and the system B2045+265 might
include a spiral galaxy (or spiral galaxies) but its lensing interpretation
remains puzzling (see \citealt{Fas99}). Excluding these three sources
we take $N_{\rm L} = 10$ for lensing solely due to the early-type galaxy
population. To be included in the likelihood as relative image separation
probabilities are B0414+054, B1030+074, J1632$-$0033, J1838$-$3427 and
B1938+666, each of which is known to be strongly lensed by a single early-type
galaxy, and B0128+437, B0739+366 and B1555+375 whose lens types are unknown.
For the last three sources we take $w_k = 0.8$ while $w_k = 1$ for the rest.
Notice that the lensed sources of the CLASS statistical sample are not
included in the likelihood of the relative image separation probabilities
because the likelihood of the CLASS statistical sample includes the relative
image separation probabilities as well as the absolute lensing probability
(see, e.g., \citealt{Cha03}). Fig.\ \ref{fig:mu-f0} shows the confidence limits
in the plane of $\mu$ (equation~\ref{fmod}) and $\sqrt{2} f_0$ 
(equation~\ref{f0star}). Here we
have chosen $\sigSIEzero = 200 \kms$ for the fiducial velocity dispersion.
Fig.\ \ref{fig:fv} shows how the factor $\sqrt{2} f (=\sqrt{2}\sigSIE/\vcvir)$
behaves as the velocity dispersion ($\sigSIE$) is varied. Notice that 
$\sqrt{2} \sigSIE$ would be equal to the circular velocity in the optical 
region if the galaxy mass profile were isothermal up to a few effective
radii. The quantity $\sqrt{2} \sigSIE$ will thus be our estimate of the 
circular velocity at the effective radius $\vceff$ from strong lensing 
based on the SIE model (equation~\ref{SIE}).
To understand better Fig.\ \ref{fig:fv} it is useful to rewrite $f$ as
\beq
f = f_0 \left( \frac{\sigSIE}{\sigSIEzero} \right)^{\mu/(\mu+1)},
\label{fsigma}
\eeq
which can be derived from equation~(\ref{fmod}) using 
equation~(\ref{f0star}). From equation~(\ref{fsigma})
we can see that for the given uncertainties of $\mu$ and $f_0$
(Fig.\ \ref{fig:mu-f0}) the uncertainty of $f$ increases as $\sigSIE$
decreases since $\mu$ is negative. We can equivalently say that the uncertainty
of the ratio $f=\sigSIE/\vcvir$ increases as $\sigSIE$ (and thus $\vcvir$)
decreases for a given uncertainty of $\vcvir$. Finally,
Fig.\ \ref{fig:sigma-vc} (Fig.\ \ref{fig:vc-sigma}) shows how $\vcvir$ 
($\sigSIE$) behaves as $\sigSIE$ ($\vcvir$) is varied.

Next we consider the case where we use only the first part in 
equation~(\ref{Lhood}),
namely the relative image separation probabilities. Notice that in this case
the absolute abundances of the early-type galaxies from the SAM have no effects
on the constraints on the velocity profiles of galaxies. For this case we use
all the lens systems used by \citet{Cha05}, namely the 8 lens systems above
along with the following 7 lens systems in the CLASS statistical sample;
 B0445+123, B0631+519, B0712+472, B1152+199, B1422+231, B1933+503, and
 B2319+051. The results for this case are also displayed in
Figures~\ref{fig:mu-f0}~to~\ref{fig:vc-sigma}.

Bearing in mind that the strict range of validity
for our analyses is $200 \kms \la \vcvir \la 600 \kms$ (see above),
 our main findings are as follows.
First, parameter $\mu$ must be negative meaning that the ratio of the inner
velocity to the virial velocity must be larger for a less massive halo
(Fig.\ \ref{fig:mu-f0}).
A constant ratio (i.e.\ $\mu = 0$) is excluded at a highly significant level.
We have $\mu = -0.78^{+0.05}_{-0.04}$ (68\% CL) based on both
the image separation distribution of radio lenses and the statistics of the
CLASS statistical sample or $\mu = -0.67^{+0.11}_{-0.08}$ (68\% CL)
based only on the image separation distribution of radio lenses. Second, 
the inner velocity dispersion $\sigSIE$ and the virial circular velocity
$\vcvir$ scale as
\beq
\frac{\sigSIE}{200 \kms} = \frac{\eta \sigcent}{200\kms} =
\left(1.17^{+0.40}_{-0.26} 
 \frac{\vcvir}{200\kms}\right)^{0.22^{+0.05}_{-0.04}}
\label{ReISDCLASS}
\eeq
for the case of using both the image separation distribution and CLASS
statistics, or
\beq
\frac{\sigSIE}{200\kms} = \frac{\eta \sigcent}{200\kms} = 
 \left( 0.82^{+0.54}_{-0.18}
 \frac{\vcvir}{200\kms} \right)^{0.33^{+0.11}_{-0.08}}
\label{ReISD}
\eeq
for the case of using only the image separation distribution. We take
\beq
\eta = \frac{\sigSIE}{\sigcent}  = 1.0 \pm 0.1
\label{VDratio}
\eeq 
from \citet{Cha05} and \citet{TK04}.
Third, for galaxies with velocity dispersion $\sigSIE \la 210$-$230\kms$
the ratio $\sqrt{2} \sigSIE / \vcvir (=\vceff/\vcvir)$ becomes increasingly 
larger than 1 as $\sigSIE$ decreases (Fig.\ \ref{fig:fv}). 
For a typical bright galaxy with $\sigSIEzero = 200 \kms$,
$\sqrt{2} f_0 = \sqrt{2} \sigSIEzero / \vcvirzero = 1.65^{+0.57}_{-0.37}$
(image separation distribution + CLASS statistics) or $1.16^{+0.76}_{-0.30}$
(image separation distribution only) at the 68\% confidence level.
However, for large galaxies with $\sigSIE \ga 210$-$230 \kms$, the ratio
$\sqrt{2} \sigSIE / \vcvir \la 1$. 
For a galaxy with $\sigSIE = 260 \kms$, which
corresponds approximately to $\vcvir \sim 550 \kms$
(see Fig.\ \ref{fig:sigma-vc} and Fig.\ \ref{fig:vc-sigma}),
$\sqrt{2} f = \sqrt{2} \sigSIE / \vcvir = 0.65^{+0.15}_{-0.12}$
(image separation distribution + CLASS statistics) or $0.68^{+0.21}_{-0.11}$
(image separation distribution only) at the 68\% confidence level.
 We discuss appropriate interpretations of these results along with
possible sources of systematic errors in \S 4.

\section{Discussion}

In this work we have compared the velocity dispersion at about the effective
radius of the optical region ($\sigSIE$) derived from strong lensing based on
the SIE lens model (equation~\ref{SIE}) with the circular velocity at the 
virial radius of the surrounding halo ($\vcvir$) predicted by semi-analytical
studies of galaxy formation for the early-type galaxy population. 
Assuming that the inner velocity dispersion is related to the virial circular
velocity by the simple power-law model (equation~\ref{fmod}),
we find that the ratio of the velocity dispersion to the circular velocity
becomes increasingly larger as the velocity dispersion decreases; the power-law
index $\mu$ in equation~(\ref{fmod}) must be negative ({Fig.}~\ref{fig:mu-f0}).
Hence, we can conclude that there is clearly the trend that the smaller the
surrounding halo is, the more enhanced the optical velocity dispersion is.
This is the most robust result from our work.
In addition to this, lensing studies (\citealt{Koc94, TK04, Cha05}) have found
 that the velocity dispersion in the optical region
derived from strong lensing based on singular isothermal ellipsoids, as is the
case in this work, is about the same as the central stellar velocity
dispersion determined from spectroscopic observations, 
namely $\eta = \sigSIE/\sigcent \approx 1$ (equation~\ref{VDratio}).

Strong lensing gives the scaling between the central velocity dispersion 
$\sigcent$ and the virial circular velocity $\vcvir$ as
$\sigcent \propto \vcvir^{0.22^{+0.05}_{-0.04}}$ (image separation distribution
+ CLASS statistics) or $\sigcent \propto \vcvir^{0.33^{+0.11}_{-0.08}}$
(image separation distribution only). Our results are broadly consistent
with those from recent other studies based on halo occupation statistics
(\citealt{VO04, Yang05}) and galaxy-galaxy weak lensing (\citealt{Man06}).
These studies all obtained some relations between the virial mass of a halo
$\Mvir$ and the luminosity $L_c$ of a central galaxy  hosted by the halo.
Suppose $L_c \propto \Mvir^{1/\gamma}$ and the Faber-Jackson relation
(\citealt{FJ76}) $L_c \propto \sigma^{\beta}$ where we shall take $\beta = 4$.
It follows then $\sigcent \propto \vcvir^{3/(\beta \gamma)}$ as
$\Mvir \propto \vcvir^3$ (see \citealt{Bul01}).
\citet{VO04} found $\gamma \approx 3.57$ using halo occupation statistics
 for massive halos which implies $\sigcent \propto \vcvir^{0.21}$.
  Using galaxy groups in the Two Degree Field Galaxy
Redshift Survey, \citet{Yang05}
found $\gamma \approx 4$ for $\Mvir \ga 10^{13}h^{-1} M_\odot$ but
$\gamma \approx 3/2$ for $\Mvir \la 10^{13}h^{-1} M_\odot$. The range of the
halo mass corresponding to our velocity range extends across the transition
mass $10^{13}h^{-1} M_\odot$, so the implied scaling will be somewhere between
$\sigcent \propto \vcvir^{0.19}$ and $\sigcent \propto \vcvir^{0.5}$.
The galaxy-galaxy weak lensing study of \citet{Man06} found
$\gamma \approx 2.7$ for galaxies with $L_c \ga L_*$ which implies
$\sigcent \propto \vcvir^{0.28}$. All these results are clearly consistent with
our results within the statistical errors.

\begin{figure*}
\begin{center}
\setlength{\unitlength}{1cm}
\begin{picture}(11,11)(0,0)
\put(-4.5,13.){\includegraphics{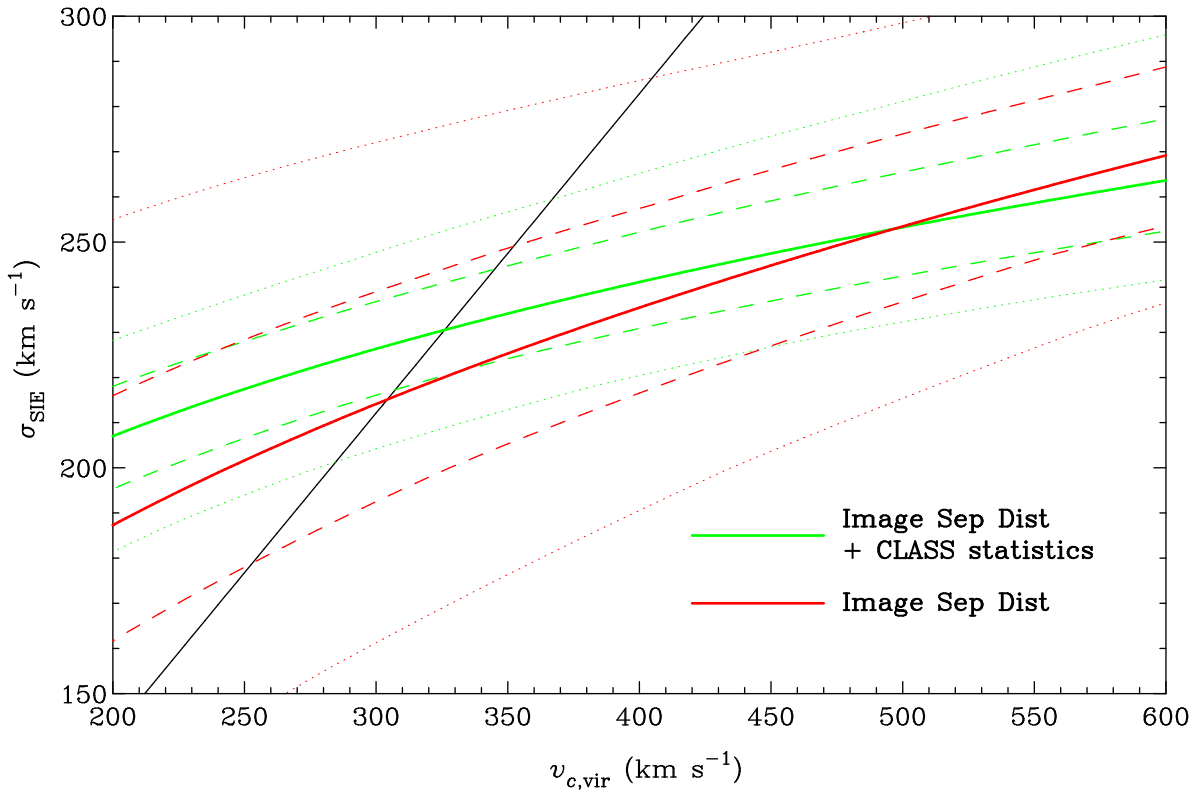}}
\end{picture}
\caption{
The behaviour of $\sigSIE$ as a function of $\vcvir$. Notations are the same as
in Fig.\ \ref{fig:sigma-vc}.
}
\label{fig:vc-sigma}
\end{center}
\end{figure*}

Our plot ({Fig.}~\ref{fig:fv}) of the ratio 
$\sqrt{2}f (=\sqrt{2} \sigSIE/\vcvir =\vceff/\vcvir)$ (equation~\ref{fdef}; 
where $\sigSIE$ is the velocity dispersion at about the effective radius 
and $\vcvir$ is the circular velocity at the virial radius) against
the velocity dispersion $\sigSIE$ shows some interesting features.
First of all, for galaxies with velocity dispersion
$\sigSIE \la 210$-$230\kms$ the ratio $\sqrt{2}f  > 1$. For example,
a typical bright galaxy with $\sigSIEzero = 200 \kms$ has
 $\sqrt{2} f_0 = 1.65^{+0.57}_{-0.37}$
(image separation distribution + CLASS statistics) or
$1.16^{+0.76}_{-0.30}$ (image separation distribution only).
Given $\vceff=\sqrt{2} \sigma$ for our lens model (equation~\ref{SIE}), 
the above results are in agreement with the \citet{Sel02} galaxy-galaxy weak 
lensing result of $\vcopt/\vcvir = 1.68 \pm 0.2$ for
an $L_*$ early-type galaxy with his adopted value of
$\sigcent = 177 \kms$.\footnote{It appears that the value of
$\vcopt/\vcvir$ changes little between $\sigcent = 177 \kms$ and
 $200 \kms$ according to the results by \citet{Sel02}.}
A striking feature of Fig.\ \ref{fig:fv} is its behaviour at high velocity
dispersions. The ratio $\sqrt{2} f$ is about 1 or less for
$\sigSIE \ga 210$-$230 \kms$. For a galaxy with $\sigSIE = 260 \kms$,
which is hosted by a halo of $\vcvir \sim 550 \kms$ (see
Fig.\ \ref{fig:vc-sigma}), our result is $\sqrt{2} f  = 0.65^{+0.15}_{-0.12}$
(image separation distribution + CLASS statistics) or $0.68^{+0.21}_{-0.11}$
(image separation distribution only) at 68\% confidence level.
The 95\% confidence range is $0.42  \leq \sqrt{2} f \leq 0.99$
(image separation distribution + CLASS statistics)  or
$0.48  \leq \sqrt{2} f \leq 1.63$ (image separation distribution only)
for $\sigSIE = 260 \kms$. From \citet{Sel02}
$\sqrt{2} f = 1.3 \pm 0.2$ for $\sigcent = 290 \kms$. 
Given that our derived velocity dispersion $\sigSIE$ is believed to be 
similar to the central stellar velocity dispersion $\sigcent$ 
(equation~\ref{VDratio}), 
our results and the  \citet{Sel02} results show some intriguing
difference although the 95\% statistical errors partially overlap.
For the large circular velocity systems ($\vcvir \sim 550\kms$), it is
likely that the haloes may host several galaxies and our inner
velocity dispersion is appropriate for a central galaxy. One possible
interpretation of our results is then that the optical galaxy is well within
the peak radius $r_{\rm peak}$ where the circular velocity is highest and the 
baryonic boost of the velocity within the optical galaxy is relatively small 
for the halo with $\vcvir \sim 550\kms$ (see below for further discussion).

While our results appear to be in reasonable agreement with halo
occupation number studies of \citet{VO04} and \citet{Yang05} and
the most recent weak lensing studies of \citet{Man06}, it is still important
to consider possible sources of systematic errors for our results.
First of all, the abundances of early-type galaxies at low
virial circular velocities from the SAM are uncertain at present. As can be
seen in Fig.\ \ref{fig:cvf}, the abundances of early-type galaxies from the
SAM appear to be abnormally high at low virial circular velocities and do not
fit into our model CVF. In fact, \citet{Kan05} find that their predicted
luminosity function of early-type galaxies does not match well observed
luminosity functions at the low end of luminosity.
Semi-analytical methods have their limitations primarily due to uncertainties
in the star formation and feedback processes (which are also present in
hydrodynamical simulations). In most semi-analytical studies it appears that
the number of galaxies at the faint end tends to increase faster than the
observed trend as the luminosity decreases. We have chosen to ignore
the lowest four data points of the SAM (Table~2) in fitting our model CVF.
As far as our analyses are limited to the range
$200 \kms \la \vcvir \la 600 \kms$, ignoring the lowest four data
points would not affect the results based on the image separation distribution
only. However, ignoring the lowest four data points may cause a systematic
error for the results based on both the image separation distribution and
CLASS statistics because the lensing rate depends on the entire range of
the velocity dispersion function but discounting the four data points may
have biased the behaviour of the circular velocity function so that the
relation between the two may be affected. Nevertheless,
it is likely that the systematic error might not be too large for
the following reasons. Our procedure does not entirely ignore the
low circular velocity galaxies but uses the extrapolated abundances
from the fitted function rather than the given abundances. Moreover,
current observations show that as the luminosity of the early-type galaxy
decreases below $L_*$ the surface brightness distribution becomes less
concentrated so that it is less effective for strong lensing.

Secondly, the abundances of early-type galaxies at the highest velocity end
might also be systematically biased. In fact, semi-analytical models tend to
over-predict the number of very luminous galaxies, although the inclusion 
of AGN feedbacks seems to cure this problem (\citealt{Bow06,Cro06}). 
An over-predicted abundance of galaxies at a given
virial circular velocity will lead to an underestimate of $f$ at the
corresponding velocity dispersion. To see this effect we have done the
following numerical experiment. We decreased the numbers of early-type
galaxies for bins $i = 9$ to 15 in Table~1 (with $\vcvir>312.5\kms$) by
successively larger proportions
from $i=9$ to 15 so that at $i=15$ the adjusted number becomes one half of the
unadjusted number. In this case we indeed find that the value of $f$ somewhat
increases at large velocity dispersions in particular at $\sigSIE = 260 \kms$
but within the 68\% statistical error.

Thirdly and finally, our results might be biased because of the assumed lens 
model, namely the SIE (equation~\ref{SIE}). Theoretically, the circular 
velocity (or the velocity dispersion) derived from strong lensing depends on 
the radial profile and the shape of the model mass distribution. 
For the observed image size and morphology, the predicted size of the Einstein 
ring $R_{\rm Ein}$ and the projected mass within the ring  $M_{\rm Ein}$ 
depend on the radial profile. However, for most of the known lens systems 
detailed mass modelling shows that the derived $R_{\rm Ein}$ and $M_{\rm Ein}$ 
vary little as the radial profile is varied. The circular velocity at 
$R_{\rm Ein}$ depends on the mass within the sphere of radius $R_{\rm Ein}$, 
$M(R_{\rm Ein})$ rather than $M_{\rm Ein}$. For a spherical mass distribution
of the form $\rho \sim r^{-\nu}$ a model with $\nu=1.7$ ($\nu=2.3$) would give
a circular velocity at $r=R_{\rm Ein}$ 12\% lower (9\% higher) than the 
isothermal model with $\nu=2$ for the same $M_{\rm Ein}$. It is possible that
the average mass profile of early-type galaxies systematically varies from 
smaller systems to larger systems. In fact, our own results appear to imply
varying structures of early-type galaxies (see below). Nevertheless, 
the errors for $f$ arising from approximating all the galaxies using isothermal
 models appear to be within the current 68\% statistical errors. 
Furthermore, the inner mass profiles of early-type galaxies at least up to the
effective radii appear to be close to isothermal from stellar dynamical 
modelling (e.g., \citealt{Rix97,Tho05}) as well as from detailed modelling of 
individual lenses (e.g., \citealt{RK05,Koo06}). Thus, the assumption of the 
SIE model might not actually cause as large errors as estimated above.
 The shape of the mass distribution also affects the derived circular 
velocity (or velocity dispersion) from strong lensing. For example, prolate and
oblate isothermal models give different results (see \citealt{Cha03}). 
In this work we have assumed that one half of galaxies are oblate and the 
other half are prolate. The average shape of early-type 
galaxies might systematically vary from smaller systems to larger systems. 
However, according to the model of \citet{Cha03} the change of the velocity
dispersion from the half oblate and half prolate case is $\sim 10\%$ even for
the extreme cases of all oblate and all prolate. 
Considering the above estimates of possible systematic errors, we can 
tentatively conclude that the essential trend of $f$ for 
$200 \kms \la \vcvir \la 600 \kms$ and the result $\sqrt{2} f \la 1$ at 
$\sigSIE = 260 \kms$ are likely to be real.

The fact that $\sqrt{2} f(=\vceff/\vcvir)$ is not unity shows that galaxies are
 not well approximated by isothermal profiles for significant parts of the
virial radii. The trend of $f$ as a function of the velocity dispersion 
constrained from strong lensing in this work is in agreement with theoretical 
expectations. First, more massive haloes have smaller concentrations 
(see \citealt{NFW97,Bul01}), and so they have smaller peak velocities relative
to the virial circular velocity compared with smaller haloes.
Second, the baryonic modifications to the inner mass profiles of the haloes
are less pronounced for more massive haloes and consequently the baryonic
boost of the velocity may be smaller. On the other hand, smaller
circular velocity systems may have a larger value of $f$ as baryonic cooling
could be more efficient in such systems (although feedback processes may
suppress cooling in very small systems). Third, for massive haloes the radius
of the optical galaxy may be much smaller than the peak radius $r_{\rm peak}$
of the halo where the circular velocity obtains its maximum while the optical
radius is comparable to $r_{\rm peak}$ for small haloes.
However, to assess the full significance of our results,
detailed modelling is necessary incorporating theretical halo profiles, 
baryonic modifications of the haloes, observed light distributions and 
realistic distribution functions (see, e.g., \citealt{vdM00} for modelling of
 galaxy  clusters). 

\begin{figure*}
\begin{center}
\setlength{\unitlength}{1cm}
\begin{picture}(14,14)(0,0)
\put(-3.,14.){\includegraphics{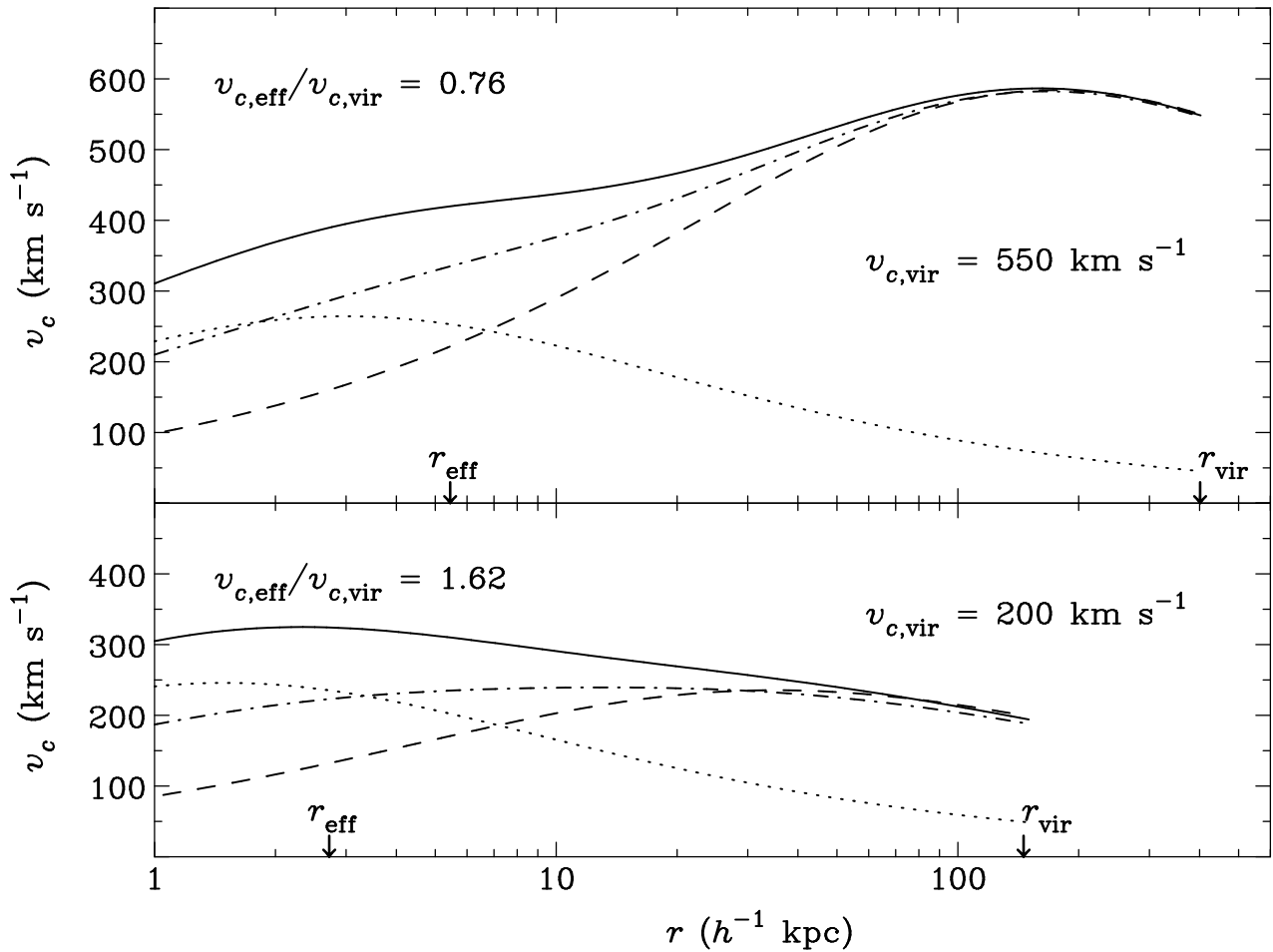}}
\end{picture}
\caption{
The circular velocity profiles for a two-component early-type galaxy model 
at $z=0.6$ consisting of a Hernquist light (stellar mass) distribution residing
in a halo, which is supposed to form from an initial NFW halo by baryonic 
cooling. The solid, dashed, dash-dotted, and dotted lines represent 
respectively the profiles for the final total mass, initial NFW halo, final 
modified halo, and Hernquist stellar mass distributions. For the lower panel 
($\vcvir=200\kms$), we take $\cvir=9$ and $\Fcool=0.09$.
For the upper panel ($\vcvir=550\kms$), we take $\cvir=5$ and 
$\Fcool=0.01$. Here $\cvir$ is the halo concentration 
(\citealt{Bul01}) and $\Fcool$ is the fraction of cooled baryons forming
stars out of the initial total mass. In these models, the predicted ratios of
the circular velocities at the effective and the virial radii 
$\vceff/\vcvir$ are consistent with our derived values of
$\sqrt{2} \sigSIE/\vcvir$.
}
\label{fig:galmod}
\end{center}
\end{figure*}

In this work, we consider only a simple two-component model recently
used in the literature (\citealt{Kee01,Sel02}) to see whether our derived
ratio of the SIE velocity dispersion to the virial circular velocity
 can be matched qualitatively and shed new light on galactic structures. 
(We postpone a  more detailed work to a future publication.) 
The model starts from an initial NFW halo and then turns into a Hernquist 
(\citealt{Her90}) light (stellar mass) distribution by baryonic infall 
residing in a modified halo. We use the adiabatic contraction model by
\citet{Blu86} to calculate the modified halo profile. Fig.~\ref{fig:galmod}
shows the constructed models at redshift $z=0.6$ for the haloes of 
$\vcvir=200\kms$ and $\vcvir=550\kms$. For the $\vcvir=200\kms$ halo, we take 
$\cvir=9$ (where $\cvir$ is the halo concentration) and  
$\Fcool = 0.09$ (where $\Fcool$ is the the cooled baryon fraction
out of the total mass).   For the $\vcvir=550\kms$ halo,  $\cvir=5$ 
and $\Fcool = 0.01$. Here we have (approximately) estimated the values 
of $\cvir$ using the simulation results of \citet{Bul01} while the values of 
$\Fcool$ have been adjusted to produce circular velocity functions that are 
consistent with the results from this work. Consequently, in these models
the predicted values of the ratio of the circular velocities at the effective 
and the virial radii $\vceff/\vcvir$ are in agreement with our derived values 
of $\sqrt{2} f$ for $\sigSIE=200\kms$ and $260\kms$. The circular velocity 
curves shown in Fig.~\ref{fig:galmod} imply evidently the following for
galaxy formation processes and resultant mass profiles. First, the galaxies
with $\vcvir=200\kms$ (and corresponding $\sigSIE=200\kms$) and  
$\vcvir=550\kms$ ($\sigSIE=260\kms$) require very different values of 
$\Fcool$ in order to be consistent with our derived values of 
$\sqrt{2}f$. The fitted values of $\Fcool$ imply that the fraction of 
infalling baryons forming stars and consequential modification to the halo
are becoming increasingly smaller as the halo gets more and more massive. 
This is qualitatively consistent with the fact that for galaxy clusters 
(the high mass limit of haloes) the NFW halo profile is well preserved.
 Second, the system of $\vcvir=200\kms$ has a density profile 
close to isothermal up to a few effective radii but
steeper than isothermal beyond. On the other hand, the system of
$\vcvir=550\kms$ has a profile shallower than isothermal up to about a half 
of the virial radius. However, more detailed investigations along with more 
precise interpretations will be considered in a future publication.

The parameterisation we use in equation~(\ref{fmod}) might not be optimal.
The current CLASS and PANELS samples are, however, too small to
allow us to explore a more realistic functional form.
However, one advantage of the method is worth emphasizing:
the lensing probability scales as $f^4$ and the separation
scales as $f^2$ -- the lensing properties therefore
depends on quite high powers of $f$. This also implies that even
if the semi-analytical modelling is somewhat uncertain, we may
still be able to put strong limits on this parameter
with the next-generation lens surveys. Current generation
hydrodynamical simulations cannot yet resolve and simulate the inner
parts of early-type galaxies realistically. For example, the high
resolution simulation of an early-type galaxy by \citet{Mez03}
predicts a central velocity dispersion as high as $650\kms$ due
to the compact size of their stellar component. So gravitational
lensing can play an important role in empirically assessing
the roles of baryonic cooling in galaxy formation and evolution.

\vspace{1cm}
We thank Yipeng Jing and Ian Browne for helpful discussions. We also thank the
anonymous referee for insightful comments that greatly improved the 
presentation and discussion. KHC acknowledges support from the Astrophysical 
Research Center for the Structure and Evolution of the Cosmos by KOSEF. 
SM was partly supported in travel by the Chinese
Academy of Sciences and the European Community's Sixth
Framework Marie Curie Research Training Network Programme,
Contract No. MRTN-CT-2004-505183 ``ANGLES''. KX acknowledges a
fellowship from the Royal Society and travel support from Jodrell Bank.

\bibliographystyle{mn2e}

\begin{thebibliography}{}

\bibitem[\protect\citeauthoryear{Blumenthal et al.} {1986}]{Blu86}
Blumenthal G. R., Faber S. M., Flores R., Primack J. R., 1986, ApJ, 301, 27

\bibitem[\protect\citeauthoryear{Bond et al.} {1991}]{Bon91}
Bond J.~R., Cole S., Efstathiou G., Kaiser N., 1991, ApJ, 379, 440

\bibitem[\protect\citeauthoryear{Bower et al.} {2006}]{Bow06}
Bower R. G., Benson A. J., Malbon R., Helly J. C., Frenk C. S., Baugh C. M.,
 Cole S., Lacey C. G., 2006, MNRAS, 370, 645

\bibitem[\protect\citeauthoryear{Bullock et al.} {2001}]{Bul01}
Bullock J.~S., Kolatt T.~S., Sigad Y., Somerville R.~S., Kravtsov A.~V.,
Klypin A.~A., Primack J.~R., Dekel A., 2001, MNRAS, 321, 559

\bibitem[\protect\citeauthoryear{Browne et al.} {2003}]{Bro03}
Browne I.~W.~A. et al., 2003, MNRAS, 341, 13

\bibitem[\protect\citeauthoryear{Chae} {2003}]{Cha03}
Chae K.-H., 2003, MNRAS, 346, 746

\bibitem[\protect\citeauthoryear{Chae} {2005}]{Cha05}
Chae K.-H., 2005, ApJ, 630, 764

\bibitem[\protect\citeauthoryear{Chae et al.} {2002}]{Cha02}
Chae K.-H. et al., 2002, Phys.\ Rev.\ Lett., 89, 151301

\bibitem[\protect\citeauthoryear{Chae \& Mao} {2003}]{CM03}
Chae K.-H., Mao S., 2003, ApJ, 599, L61

\bibitem[\protect\citeauthoryear{Cole et al.} {2000}]{Col00}
Cole S., Lacey C.~G., Baugh C.~M., Frenk C.~S., 2000, MNRAS, 319, 168

\bibitem[\protect\citeauthoryear{Comerford et al.} {2006}]{Com06}
Comerford J.~M., Meneghetti M., Bartelmann M., Schirmer M., 2006,
ApJ, 642, 39

\bibitem[\protect\citeauthoryear{Croton et al.} {2006}]{Cro06}
Croton D.~J., Springel V., White S.~D.~M., de Lucia G., Frenk C.~S., Gao L.,
 Jenkins A., Kauffmann G., Navarro J.~F., Yoshida N., 2006, MNRAS, 365, 11

\bibitem[\protect\citeauthoryear{de Blok} {2005}]{deB05}
de Blok W.~J.~G., 2005, ApJ, 634, 227

\bibitem[\protect\citeauthoryear{Faber \& Jackson}{1976}]{FJ76}
Faber S.~M., Jackson R.~E., 1976, ApJ, 204, 668

\bibitem[\protect\citeauthoryear{Fassnacht al.} {1999}]{Fas99}
Fassnacht C. D. et al., 1999, AJ, 117, 658

\bibitem[\protect\citeauthoryear{Fukugita et al.} {1992}]{Fuk92}
 Fukugita M., Futamase T., Kasai M., Turner E.~L., 1992, ApJ, 393, 3

\bibitem[\protect\citeauthoryear{Helbig et al.} {1999}]{Hel99}
 Helbig P., Marlow D., Quast R., Wilkinson P.~N., Browne I.~W.~A.,
 Koopmans L.~V.~E., 1999, A{\&}AS, 136, 297

\bibitem[\protect\citeauthoryear{Hernquist} {1990}]{Her90}
 Hernquist L., 1990, ApJ, 356, 359

\bibitem[\protect\citeauthoryear{Jing \& Suto}{2002}]{JS02}
Jing Y.~P.,  Suto Y., 2002, ApJ, 574, 538

\bibitem[\protect\citeauthoryear{Kang et al.} {2005}]{Kan05}
 Kang X., Jing Y.~P., Mo H.~J., B\"orner G., 2005, ApJ, 631, 21

\bibitem[\protect\citeauthoryear{Kauffmann et al.} {1999}]{Kau99}
Kauffmann G., Colberg J.~M., Diaferio A., White S.~D.~M., 1999, MNRAS, 307, 529

\bibitem[\protect\citeauthoryear{Keeton} {2001}]{Kee01}
 Keeton C.~R., 2001, ApJ, 561, 46

\bibitem[\protect\citeauthoryear{Kochanek} {1994}]{Koc94}
 Kochanek C.~S., 1994, ApJ, 436, 56

\bibitem[\protect\citeauthoryear{Kochanek} {1995}]{Koc95}
 Kochanek C.~S., 1995, ApJ, 453, 545

\bibitem[\protect\citeauthoryear{Kochanek} {1996}]{Koc96}
 Kochanek C.~S., 1996, ApJ, 466, 638

\bibitem[\protect\citeauthoryear{Koopmans et al.} {2006}]{Koo06}
 Koopmans L. V. E., Treu T., Bolton A. S., Burles S., Moustakas L. A., 2006,
 ApJ, 649, 599

\bibitem[\protect\citeauthoryear{Lacey \& Cole} {1994}]{LC94} 
 Lacey C., Cole S., 1994, MNRAS, 271, 676

\bibitem[\protect\citeauthoryear{Mandelbaum et al.} {2006}]{Man06}
 Mandelbaum R., Seljak U., Kauffmann G., Hirata C.~M., Brinkmann J.,
 2006, MNRAS, 368, 715

\bibitem[\protect\citeauthoryear{Mao} {1991}]{Mao91}
 Mao S., 1991, ApJ, 380, 9
       
\bibitem[\protect\citeauthoryear{Mao \& Kochanek} {1994}]{MK94} 
 Mao S., Kochanek C. S., 1994, MNRAS, 268, 569

\bibitem[\protect\citeauthoryear{Marzke et al.} {1998}]{Mar98}
 Marzke R.~O., da Costa L.~N., Pellegrini P.~S., Willmer C.~N.~A.,
 Geller M.~J., 1998, ApJ, 503, 617

\bibitem[\protect\citeauthoryear{Meza et al.}{2003}]{Mez03}
Meza A., Navarro J.~F., Steinmetz M., Eke V.~R., 2003, ApJ, 590, 619

\bibitem[\protect\citeauthoryear{Mitchell et al.} {2005}]{Mit05}
 Mitchell J.~L., Keeton C.~R., Frieman J.~A., Sheth R.~K., 2005, ApJ,
 622, 81

\bibitem[\protect\citeauthoryear{Myers et al.} {2003}]{Mye03}
 Myers S.~T. et al., 2003, MNRAS, 341, 1

\bibitem[\protect\citeauthoryear{Narayan \& White}{1988}]{NW88}
 Narayan  R., White S.~D.~M., 1988, MNRAS, 231, 97

\bibitem[\protect\citeauthoryear{Navarro et al.}{1997}]{NFW97}
Navarro J.~F., Frenk C.~S., White S.~D.~M., 1997, ApJ, 490, 493

\bibitem[\protect\citeauthoryear{Navarro et al.}{2004}]{Nav04}
Navarro J.~F., Hayashi E., Power C., Jenkins A.~R., Frenk C.~S.,
White S.~D.~M., Springel V., Stadel J., Quinn T.~R., 2004, MNRAS, 349, 1039

\bibitem[\protect\citeauthoryear{Ofek et al.}{2003}]{Ofe03}
Ofek E.~O., Rix H.-W., Maoz D., 2003, MNRAS, 343, 639

\bibitem[\protect\citeauthoryear{Peacock et al.}{2001}]{Pea01}
 Peacock J.~A. et al., 2001, Nature, 410, 169

\bibitem[\protect\citeauthoryear{Press \& Schechter}{1974}]{PS74}
 Press W.~H., Schechter P., 1974, ApJ, 187, 425

\bibitem[\protect\citeauthoryear{Rix et al.}{1997}]{Rix97}
Rix H.-W., de Zeeuw P. T., Cretton N., van der Marel R. P., Carollo C. M.,
1997, ApJ, 488, 702

\bibitem[\protect\citeauthoryear{Rusin \& Kochanek}{2005}]{RK05}
 Rusin D., Kochanek C.~S., 2005, ApJ, 623, 666

\bibitem[\protect\citeauthoryear{Schechter}{1976}]{Sch76}
 Schechter P., 1976, ApJ, 203, 297

\bibitem[\protect\citeauthoryear{Seljak} {2002}]{Sel02}
 Seljak U., 2002, MNRAS, 334, 797

\bibitem[\protect\citeauthoryear{Sheth \& Tormen} {2002}]{ST02}
  Sheth R.~K., Tormen G., 2002, MNRAS, 329, 61

\bibitem[\protect\citeauthoryear{Sheth et al.} {2003}]{She03}
 Sheth R.~K. et al., 2003, ApJ, 594, 225

\bibitem[\protect\citeauthoryear{Simien \& de Vaucouleurs}{1986}]{Sd86}
 Simien F., de Vaucouleurs G., 1986, ApJ, 302, 564

\bibitem[\protect\citeauthoryear{Somerville \& Primack}{1999}]{SP99}
 Somerville R.~S., Primack J.~R., 1999, MNRAS, 310, 1087

\bibitem[\protect\citeauthoryear{Spergel et al.} {2003}]{Spe03}
 Spergel D.~N. et al., 2003, ApJS, 148, 175

\bibitem[\protect\citeauthoryear{Spergel et al.} {2006}]{Spe06}
 Spergel D.~N. et al., 2006, preprint (astro-ph/0603449)

\bibitem[\protect\citeauthoryear{Springel \& Hernquist}{2003}]{SH03}
Springel V., Hernquist L., 2003, MNRAS, 339, 289

\bibitem[\protect\citeauthoryear{Swaters et al.} {2003}]{Swa03}
Swaters R.~A., Madore B.~F., van den Bosch F.~C., Balcells M.,
2003, ApJ, 583, 732

\bibitem[\protect\citeauthoryear{Tegmark et al.} {2004}]{Teg04}
 Tegmark M. et al, 2004, ApJ, 606, 702

\bibitem[\protect\citeauthoryear{Thomas et al.} {2005}]{Tho05}
Thomas J., Saglia R. P., Bender R., Thomas D., Gebhardt K., Magorrian J.,
 Corsini E. M., Wegner G., 2005, MNRAS, 360, 1355

\bibitem[\protect\citeauthoryear{Treu \& Koopmans} {2004}]{TK04}
 Treu T., Koopmans L.~V.~E., 2004, ApJ, 611, 739

\bibitem[\protect\citeauthoryear{Turner et al.} {1984}]{TOG}
Turner E.~L., Ostriker J.~P., Gott J.~R., III, 1984, ApJ, 284, 1

\bibitem[\protect\citeauthoryear{Vale \& Ostriker} {2004}]{VO04}
Vale A., Ostriker J.~P., 2004, 353, 189

\bibitem[\protect\citeauthoryear{van der Marel et al.} {2000}]{vdM00}
van der Marel R.~P., Magorrian J., Carlberg R.~G., Yee H.~K.~C.,
Ellingson E., 2000, AJ, 119, 2038

\bibitem[\protect\citeauthoryear{Voigt \& Fabian} {2006}]{VF06}
Voigt L.~M., Fabian A.~C., 2006, MNRAS, 368, 518

\bibitem[\protect\citeauthoryear{Winn et al.} {2000}]{Win00}
 Winn J.~N. et al., 2000, AJ, 120, 2868

\bibitem[\protect\citeauthoryear{Winn et al.} {2002a}]{Win02a}
 Winn J.~N., Lovell J.~E.~J., Chen H.-W., Fletcher A.~B., Hewitt J.~N.,
 Patnaik A.~R., Schechter P.~L., 2002a, ApJ, 564, 143

\bibitem[\protect\citeauthoryear{Winn et al.} {2002b}]{Win02b}
 Winn J.~N., Morgan N.~D., Hewitt J.~N., Kochanek C.~S., Lovell J.~E.~J.,
 Patnaik A.~R., Pindor B., Schechter P.~L., Schommer R.~A., 2002b, AJ,
 123, 10

\bibitem[\protect\citeauthoryear{Winn et al.} {2001a}]{Win01a}
 Winn J.~N., Hewitt J.~N., Patnaik A.~R., Schechter P.~L., Schommer R.~A.,
 L\'{o}pez S., Maza J., Wachter S., 2001a, AJ, 121, 1223

\bibitem[\protect\citeauthoryear{Winn et al.} {2001b}]{Win01b}
 Winn J.~N., Hewitt J.~N., Schechter P.~L., 2001b, in Gravitational
 Lensing: Recent Progress and Future Goals,  ASP Conference Proceedings, Vol.\
 237, ed.\ T.~G. Brainerd \& C.~S. Kochanek (San Francisco: ASP), p.\ 61

\bibitem[\protect\citeauthoryear{Yang et al.} {2005}]{Yang05}
Yang X., Mo H.~J., Jing Y.~P., van den Bosch F.~C., 2005, MNRAS,
358, 217

\end{thebibliography}

\end{document}